# Clock-centric Serial Links for the Synchronization of Distributed Readout Systems

D. Calvet

*Abstract*—Detector readout systems for medium to large scale physics experiments, and instruments in some other fields as well, are generally composed of multiple front-end digitizer boards distributed over a certain area. Often, this hardware has to be synchronized to a common reference clock with minimal skew and low jitter. Today's mainstream solutions to precise clock distribution and deterministic latency messaging rely on the capabilities of high speed serial transceivers (a.k.a. SerDes) embedded in modern Field Programmable Gate Arrays (FPGAs). An alternative option uses distinct clock and data links. This can potentially reach higher synchronization accuracy, at significant hardware expenses. This work reports some first steps to explore a third scheme for clock and synchronous message distribution. Like the standard approach, the same media is used to convey clock and data, but instead of using today's "data-centric" links where the recovered clock is only a by-product of a SerDes, this paper defines and investigates "clock-centric" links where, at the opposite, a clock is carried by the link, and synchronous data are embedded into it by a modulation technique. After defining the concepts and principles of data-centric links, experimental studies are presented. Finally, the merits and limitations of the proposed approach are discussed.

## I. Introduction

Detectors in today's medium to large scale nuclear and high energy physics experiments comprise thousands to millions of channels spread in a volume that typically ranges from around one to thousands of cubic meters. In general, all the associated readout electronics has to be synchronized to a common global reference clock (e.g. 40 MHz, 100 MHz, or some other value) with fixed stable skew, and low to ultra-low clock jitter, from 100 ps to 10 ps and even less for the most demanding future applications [1], [2]. Asynchronous, but also synchronous message diffusion with deterministic latency are commonly required capabilities for the network means deployed to synchronize distributed instrumentation.

While simple techniques for clock distribution like a multi-drop cable, an active or passive fanout (e.g. optical splitter) have been used for decades and remain adequate in some cases [3], the advent of high speed serial transceivers embedded in Field Programmable Gate Arrays (FPGAs) has greatly simplified the problem of synchronization between a master node and distributed slave nodes. The master node drives its serial transmitter(s) with the reference clock to forward, scaled-up in frequency by the required serialization factor. Each slave node recovers the serial clock from a link connected to the master and divides it to obtain a copy of the original reference clock. One difficulty is that FPGA SerDes usually include elastic buffers on the transmitter and receiver sides to decouple the user clocks at both ends from the serial link clock. Bypassing the elastic buffers at both ends – if that is possible – and locking the frequency of the serial clock of the transmitter to a multiple of the clock to forward can solve this issue. FPGA SerDes receivers never provide access to the serial clock directly. At best, only a divided version is accessible to the user. Clock frequency division introduces among receivers an unpredictable phase uncertainty, which is a multiple of the serial link Unit Interval (UI), unless the corresponding dividers can be synchronized to a fixed relative offset across all receivers, or their mutual relative phase be known. To my knowledge, the first publication that reported complete workarounds all these difficulties is [4] where deterministic latency clock forwarding was shown with the SerDes of Xilinx Virtex 5 and Spartan 6 FPGA families. A comparable achievement for Xilinx 7 series GTP transceivers was presented in [5]. WhiteRabbit [6] combines the above general concepts, and other ideas, with Ethernet technology. This technique became part of IEEE 1588 Precise Time Protocol standard in 2019 under the denomination "High Accuracy profile" [7]. Today's most advanced clock distribution solutions reach sub-UI phase determinism, down to ~1 ps, using Xilinx Ultrascale GTY/H transceivers [8]. Are the above methods definitive answers to the problem of clock and synchronous data distribution? Possibly, but some other schemes could remain attractive, even if these do not reach the ultimate performance of today's best techniques.

Over the last four decades, the increase of serial link speed has been spectacular, reaching 56 Gbps per link for today's fastest FPGAs. More and more sophisticated data encoding techniques have always had the goal to maximize the transport capacity of a link media. Clock recovery at the receiver end is primarily meant for serial data capture, and it is not designed to reproduce with the highest possible accuracy a relatively low frequency clock forwarded by the transmitter. High data throughput and moderate clock reproduction fidelity are at the opposite of what is normally needed for the synchronization network of a distributed data acquisition system: the purity of the forwarded clock can be of primary importance, while the required data bandwidth is usually modest. For example, sending 100 bit long trigger messages at 1 MHz only requires 100 Mbps of data bandwidth, which is well below the capacity of modern links. These observations made me imagine "clock-centric" links which primarily aim at transporting a reference clock with minimal degradation, while carrying user data is done less efficiently than ordinary "data-centric" links.

## II. CLOCK-CENTRIC SERIAL LINKS

A binary clock signal is a signal that features transitions from one level to the opposite level at periodical time intervals. A side effect of commonly used serial data encoding techniques is to obfuscate the carrier clock by unevenly skipping some of its transitions. Missing edges are reconstructed, more or less perfectly, by the receiver with a Clock and Data Recovery (CDR) circuit. By modulating a physical characteristic of a clock signal which is different, and preferably independent, of the property that conveys its timing content, a clock signal can transport data. Amplitude Modulation (AM), used since the birth of radio communication well illustrates this concept. This work investigates duty cycle modulation to piggyback serial data onto a reference carrier clock. The time intervals between the sensitive edges (rising or falling) of the carrier clock are entirely preserved, while the time when the non-sensitive edges occur is varied to encode user data. This simply corresponds to one of the variants of Pulse Width Modulation (PWM), a technique commonly employed for various applications, although its use for serial data transmission has long been abandoned in favor of more efficient techniques – though PWM-based serial links running at 400 Mbps and 3 Gbps have been shown in [9] and [10] respectively. Jitter analysis of PWM signals has been reported in [11]. However, I have not found that PWM has already been attempted, succeeded or failed, to perform high precision clock distribution combined with serial data transmission. I found the idea worth studying, and to distinguish it from the other usages of PWM, I named the technique "Clock Duty Cycle Modulation" (or "Modulated" when used as an adjective), abbreviated CDCM.

## III. SERIAL LINKS WITH CLOCK DUTY CYCLE MODULATION

### A. Introductory Example Based on 3-bit Serialization

Consider a 3-bit Parallel In / Serial Out (PISO) block fed, at frequency $F_0$, with 3-bit words as shown in Fig. 1. The first two bits of parallel words are set to a constant value "01" while the third bit encodes one bit of user data according to an identity conversion rule: a 0 is encoded by 0 and a 1 is encoded by 1. Setting the user data bits constantly to zero and one leads to the transmission of a clock signal at frequency $F_0$, with 33% and 66% duty cycle respectively. Sending a random flow of user data bits still produces a clock signal, but with ±16% duty cycle modulation around a 50% mean when the input is balanced. The time intervals between any two consecutive rising edges of the output stream $S_o$ are constant and equal to $T = 1/F_0$.

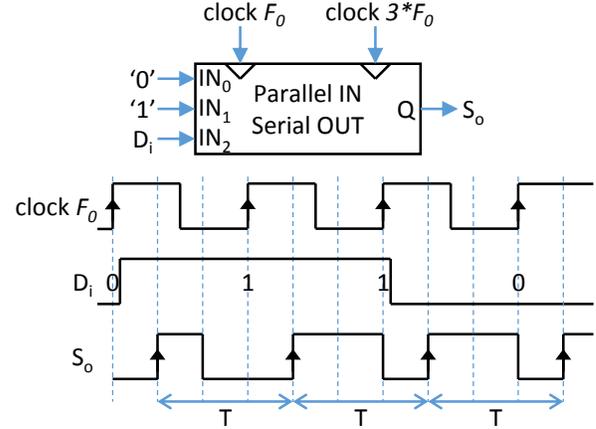

Fig. 1. Conceptual circuit to convey a clock of frequency $F_0$ over a serial link running at $3*F_0$ rate and simultaneously transport serial data at rate $F_0$.

In what follows, I call a CDCM signal "a clock" but truly speaking, it is a serial bit stream that mimics a clock. It differs from the signal that would be obtained by feeding a pure clock signal directly to a media, without serialization being involved.

### B. Generalization to N-bit Serialization

The generalization of the above encoding scheme to an N-bit serializer running at $N*F_0$ serial rate is straightforward. The N bits that compose each cycle of the transported clock contain the following slices (or Unit Intervals following the usual terminology):

- Two mandatory header bits equal to "01" (or "10" depending on the desired sensitive edge of the transported clock),
- P data payload bits that contain at only one place a transition from '1' to '0' (or '0' to '1' for a negative edge clock). These P bits may encode no information, or one to several bits of user data.

Any circular shift of the above defined N bits is also adequate. For clarity, these encoding schemes are referred to as "CDCM-N-Q", where N is the total number of UI per clock period, and Q is the average number of user data bits that can be encoded per clock cycle. Owing to the fact that the P payload bits can encode P+1 distinct combinations because these employ unary encoding, we have:

$$Q = \log_2 (P+1) \quad (1)$$

The encoded clock signal shown in Fig. 1 is referred to as a "CDCM-3-1" signal using the above denomination. The scheme corresponding to CDCM-5-2 is shown in Fig. 2.

More generally, the maximum data transport efficiency, noted $E_{max}$, of code CDCM-N-Q is reached when P is maximal, i.e. $P_{max}= N-2$:

$$E_{max} = Q_{max} / N = \log_2 (N-1) / N \quad (2)$$

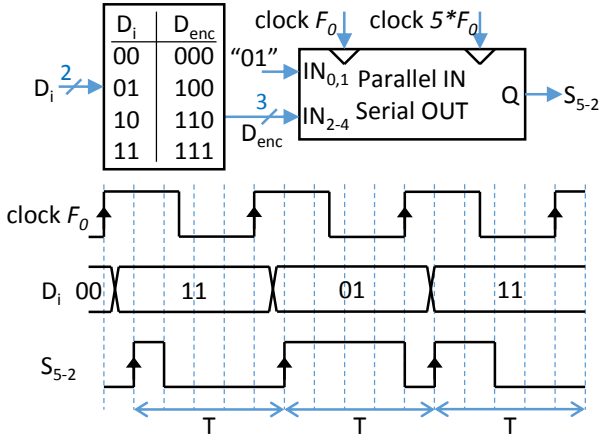

Fig. 2. Serial encoder for CDCM-5-2 scheme. A combinatorial encoder is used to convert two bits of user data into three bits of unary encoded payload prior to serialization at rate $5*F_0$. The actual user data rate is $F_0*2/5$.

The optimal coding efficiency and optimal number of user data bits that can be transported per clock cycle are plotted on Fig. 3 for serializers that have a granularity from 2 to 20 bits per carrier clock cycle. The simplest code, CDCM-3-1, can transport at most one bit of data per cycle. It has an efficiency of 33%. The peak efficiency is reached by code CDCM-5-2 which can transport 2 bits of user data per cycle, i.e. a coding efficiency of 40%. All CDCM codes are much less efficient than those commonly employed for serial data transmission, e.g. 80% for 8B/10B encoding. But recall that the technique is primarily meant to transport a clock signal with attempted superior fidelity, and conversely modest data volumes.

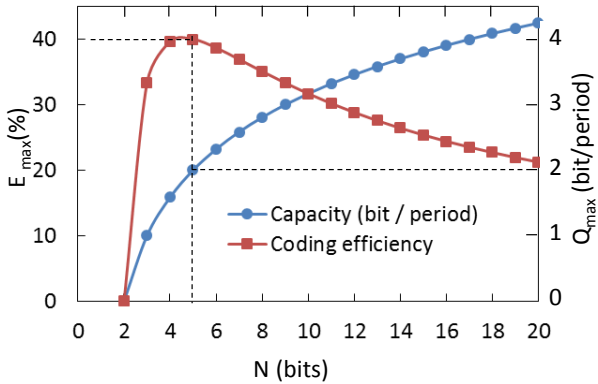

Fig. 3. Maximal data transport capacity per transmitted clock period and data transport efficiency versus the number of time slices per cycle.

### C. Exploiting Ternary Symbol Transmit Capability

An interesting code is CDCM-4-1.5 which allows the encoding of a ternary symbol per cycle. Let the data portion of P bits be "10" when no user data has to be transmitted, and "00" and "11" to transmit a 0 and a 1 respectively. As it can be seen on Fig.4, the serial stream on the link when it is idle is simply the pure carrier clock with constant 50% duty cycle. During data transmission, it becomes a 25% or 75% duty cycle clock period for a 0 and a 1 being transmitted respectively.

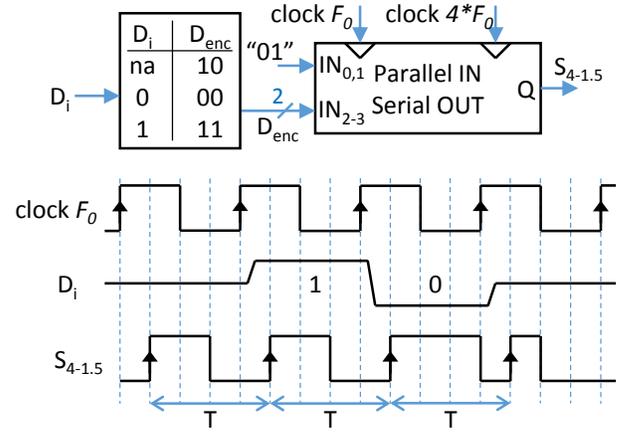

Fig. 4. Serial encoder for CDCM-4-1.5 scheme. When no user data is transmitted, $S_{4-1.5}$ is a 50% duty cycle clock at frequency $F_0$. This scheme can transport user data at a rate up to $F_0$ with a serial link that runs at $4*F_0$ bit rate.

### D. Sub-optimal Payload Filling

It should be noted that P can be chosen to be less than (N-2). This is exemplified on Fig. 5 with code CDCM-20-1.5 where the 18 payload bits of idle symbols are encoded by nine 1's and nine 0's, a transmitted 0 is coded by eight 1's and ten 0's and a transmitted 1 is coded by ten 1's followed by eight 0's. The resulting signal is a 50% duty cycle clock in the absence of data transmission. The duty cycle of this clock only changes between 45% and 55% when data are transmitted. SerDes embedded in modern mid-range FPGAs can easily reach 2 Gbps speed. Using CDCM-20-1.5 encoding, a link based on such SerDes can transport a 100 MHz clock – ten 0's and ten 1's serialized at 2 Gbps – that embeds 100 Mbps of user data in its slightly distorted duty cycle of ±5% around a mean of 50%.

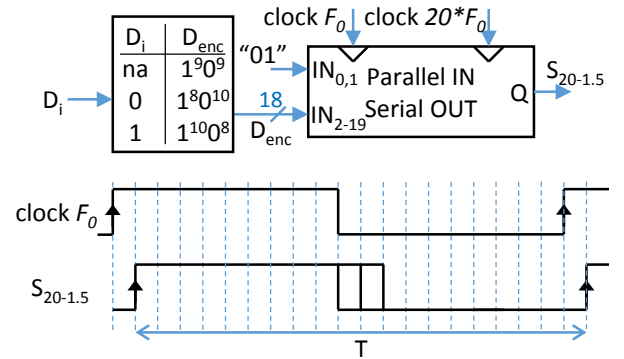

Fig. 5. Encoder to transport a clock at frequency $F_0$ and user data at bit rate $F_0$ using a serializer running at $20*F_0$. The numbers in superscript indicate how many times each bit is repeated.

### E. Specific Properties of CDCM–N–1 Schemes

For CDCM–N–1 schemes, the actual user data bit rate is identical to the frequency of the transmitted clock. Hence, neither multiplication nor division of the forwarded clock is involved for data communication. Let the encoding of the user data bit carried within each clock cycle be that given in Table I.

The following properties hold:

1) For a balanced input data stream, the modulated clock signal is DC-balanced.

2) The duty cycle distortion of every clock cycle is minimal and takes two symmetric values around 50% depending on the transmitted bit.

3) The optimal sampling time for the transported data bit is exactly ½ period after the rising edge of the forwarded clock.

TABLE I. USER DATA BIT ENCODING FOR CDCM-N-1 SCHEMES. EXPONENTS DENOTE THE NUMBER OF TIMES EACH BINARY CODING SYMBOL IS REPEATED.

|  | Encoding for a 0 | Encoding for a 1 | Encoded Clock Duty Cycle |
|---|---|---|---|
| Odd case $N = (2*k+1); k \geq 1$ | $1^{(k-1)} 0^k$ | $1^k 0^{(k-1)}$ | $0.5 \pm 1/2N$ |
| Even case $N = (2*k); k \geq 2$ | $1^{(k-2)} 0^k$ | $1^k 0^{(k-2)}$ | $0.5 \pm 1/N$ |

These properties are illustrated with CDCM-5-1 and CDCM-16-1 schemes on Fig. 6(a) and Fig. 6(b) respectively.

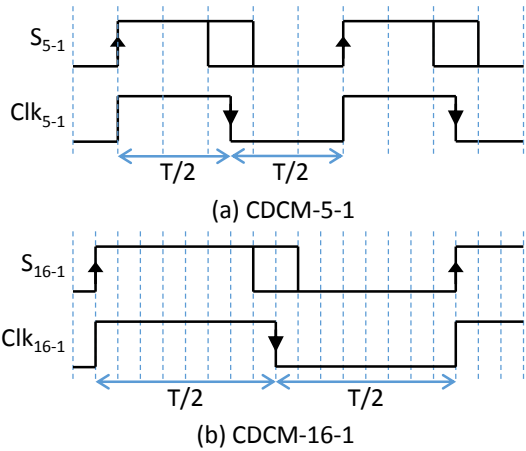

(a) CDCM-5-1

(b) CDCM-16-1

Fig. 6. Encoded signal and forwarded clock for CDCM-5-1 scheme (a) and CDCM-16-1 scheme (b). The optimal sampling point for the carried data bit is at mid-period of the transported clock. The carried bit occupies 1 UI for CDCM-N-1 scheme when N is odd, and 2 UI when N is even.

After defining the concepts of clock-centric links, the following section describes some possible structures of transmitters and receivers suited to CDCM schemes.

IV. TRANSMITTERS, RECEIVERS AND FANOUT FOR CDCM

A. Transmitters

A transmitter for CDCM can be composed of three cascaded stages. The first stage encodes user data so that the number of 0's and 1's is statistically balanced. Commonly known techniques can be used: Manchester encoding, 8B/10B encoding, scrambling [12]. The second stage is a combinatorial encoder that transforms one or several input data bits into the appropriate unary symbol representation. This stage also adds the required constant preamble and trailer bits to form the parallel word of N-bits that makes one modulated clock cycle. The last stage of the transmitter is a PISO converter driving the transport media (cable or electro-optical converter) at $N*F_0$ rate. In an FPGA, an embedded high speed SerDes can be used or a regular I/O pin that has serial transmission capability. Dedicated FPGA SerDes can reach multi-Gbps speeds but generally only support parallel word widths in multiples of 8 or 10 bits, while regular I/O pins are typically limited to ~1 Gbps but operate with more diverse parallel words, e.g. 2, 3, 4, 5, 6, 7, 8, 10 or 14 bits with Xilinx 7 series OSERDES2 primitive.

An example of a CDCM-8-1 transmitter with Manchester data pre-encoding is shown on Fig. 7. Throughput is reduced from one to 0.5 bit of user data per clock cycle because of Manchester pre-encoding. Serialization produces a clock at frequency $F_0$ composed of one 43% duty cycle period followed by one 57% duty cycle period, in an order determined by the value of the user bits $D_i$ which are supplied at frequency $F_0/2$. Despite a modest 7% coding efficiency, this scheme can be attractive for its extreme hardware simplicity (e.g. for ASIC implementation) and the minimal perturbations caused to the carrier clock.

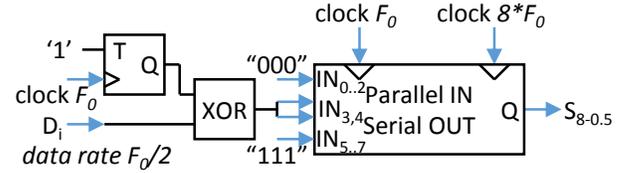

Fig. 7. Transmitter CDCM-8-1 with Manchester data pre-encoding.

B. Receivers

The typical structure of a receiver for traditionally encoded serial data is shown in Fig. 8.a. A CDR recovers from input transitions the serial clock $SER\_CLK_{REC}$ at frequency $N*F_0$ and performs re-timing of the serial data. A clock divider produces the parallel clock $PAR\_CLK$ at frequency $F_0$. The serial and parallel clocks drive a Serial In / Parallel Out (SIPO) block that converts the re-timed serial data into N-bit wide parallel data words at frequency $F_0$. This circuitry is a simplified version of the receiver part of a FPGA SerDes. Discrete CDR circuits are also available on the market, for example the ADN2814 from Analog Devices and Micrel SY87701AL.

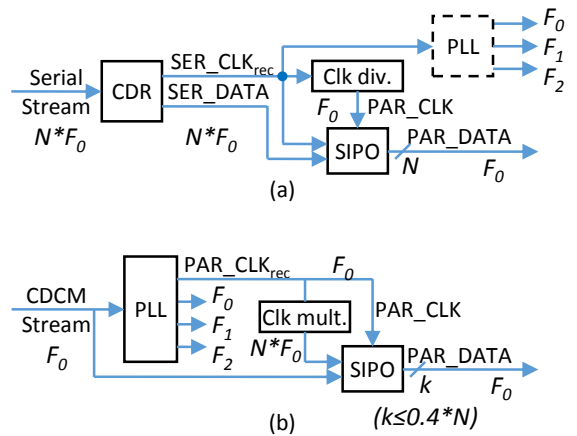

Fig. 8. Architecture of a traditional serial data receiver (a). Receiver structure for a CDCM signal (b).

Additionally, a Phase Locked Loop (PLL) can be used to attenuate the jitter of the recovered clock, or derive synchronous clocks that run at a different frequency. This well-proven structure is optimal for data transmission, but it has several imperfections for the transport of a reference clock.

Firstly, the jitter and delay properties of the CDR are usually not very well documented. Secondly, the generation of the parallel clock involves the division of the recovered clock, which is an operation that intrinsically introduces a phase-uncertainty. As earlier said, techniques have been published to overcome these difficulties, but part of the reported methods rely on features that tend to be specific to a FPGA family, or require the knowledge of technical details on the SerDes that vendors do not necessarily disclose.

The structure of a receiver for a CDCM stream is shown in Fig.8.b. Because the transported signal is effectively a clock running at frequency $F_0$, an ordinary PLL can directly purify this forwarded clock. In contrast, a CDR additionally needs to re-create all the missing transitions on encoded data to recover the serial link clock and post divide it. The sensitive edge of the phase detector of the PLL must not be the modulated edge of the incoming signal to filter out duty cycle modulation. If a zero-delay PLL is used, $PAR\_CLK_{REC}$ is phase aligned with the CDCM input, within the alignment precision of that PLL. The clock for data capture is obtained by *multiplication* of the received clock, which does not introduce phase uncertainty. The SIPO stage extracts the carried encoded data. The PLL and the clock multiplier could be implemented in the same dedicated component, they could be split between an external PLL and clocking resources within an FPGA, or both could reside within an FPGA (or an ASIC). The SIPO can easily be implemented in an FPGA (regular I/O pin or high speed SerDes). In a CDCM receiver, the carrier clock is obtained and jitter filtered directly by a PLL placed at the closest place where the signals from the physical layer of the link are available, while the traditional receiver architecture places this stage after an FPGA SerDes.

*C. Fanout*

In this section, the term "fanout" refers to an active element that regenerates the input clock and data stream before it produces one or several copies. Optionally, it can alter or enrich the original data stream. Using traditional serial data links, a fanout typically consists of a mid-range to high-end FPGA embedding multiple SerDes (one at the input and one for every output in general), an external clock purification PLL, and ancillaries. In contrast to this complex, power consuming and expensive hardware, the structure of a fanout for CDCM can be made as simple as it is shown in Fig.9. At input, the upper half of a dual fanout buffer produces the two initial copies of the input stream while the lower half provides delay compensation in the feedback loop of the PLL for improved phase matching. The narrow band PLL filters the incoming clock jitter and produces the appropriately phase-shifted and frequency multiplied clock for input stream reproduction by an array of D Flip-Flops. These may be followed by fanout buffers for increasing the number of outputs at the expense of adding jitter. All jitter sensitive components shall preferably be placed on a clean power supply island, decoupled from that of the FPGA. An entry-level FPGA (no high speed SerDes is required up to ~1 Gbps) can be added if it is desired to modify locally or differentiate on a per output basis transmitted data. Changing FPGA firmware, CDCM can be replaced by usual serial data encoding on slave ports only. The proposed technique can be tried at no risk on programmable hardware. Only a clock buffer to make two copies of the input stream has to be planned at board design. The delay incurred by a fanout for CDCM as shown in Fig. 9 is mostly determined by the phase shift set in the PLL to clock the D Flip-Flop layer which copies the input flow. In theory, the optimal value for this delay is 0.5 UI. Additional skew and jitter are incurred by the Flip-Flop layer, but this can be very small, e.g. 200 ps typical delay and 0.8 ps rms maximum additive jitter for ON Semiconductor NB7V52M 10 GHz differential D Flip-Flop.

The jitter transfer function of a fanout for CDCM is essentially determined by that of the clock regeneration PLL. Performance and stability of single and cascaded relay stages can be anticipated by simulations and PLL settings can be tuned using the model and tools normally provided by the vendor of the device. Reproducibility on hardware can be expected. It is unsure that the same can be done with FPGA embedded SerDes because loop filter settings and jitter analysis tools comparable to that of a PLL do not exist or are not publicly accessible.

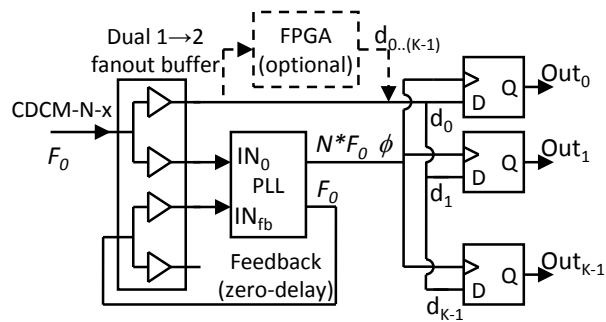

Fig. 9. Structure of a fanout for CDCM-N-X.

The following section describes the demonstrators that have been assembled to study the implementation of some CDCM schemes and investigate their performance.

V. TEST PLATFORMS

*A. Transmitter*

The transmitter used in this study is built around a TE0714 FPGA module from Trenz Gmbh [13] and its associated baseboard equipped with a Small Form factor Pluggable (SFP) optical transceiver. The module is equipped with a XC7A50T Xilinx Artix 7 FPGA. The SFP is connected to one of the four 6.6 Gbps capable transceivers of the FPGA. Using a separate electro-optical converter board, it is also possible to interface a SFP optical transceiver to regular LVDS I/O pins (680 Mbps / 1250 Mbps in SDR / DDR mode respectively). In this work, I used a GTP transmitter, running at a serial rate of 1.25 Gbps and 2.5 Gbps, clocked from an on-board oscillator. The 20 bit wide input bus runs at 62.5 MHz and 125 MHz respectively and transmits one bit of user data per clock cycle, or one bit every two cycles when Manchester pre-encoding is enabled. A typical CDCM-20-1 signal obtained with this generator is shown in Fig. 10. The carrier frequency is 125 MHz. Duty cycle is modulated by ±10% around 50% by a PRBS15 data stream at 125 Mbps. The interval between the rising edges is 8 ns. The opening of the data eye centered at half clock period is 1.6 ns.

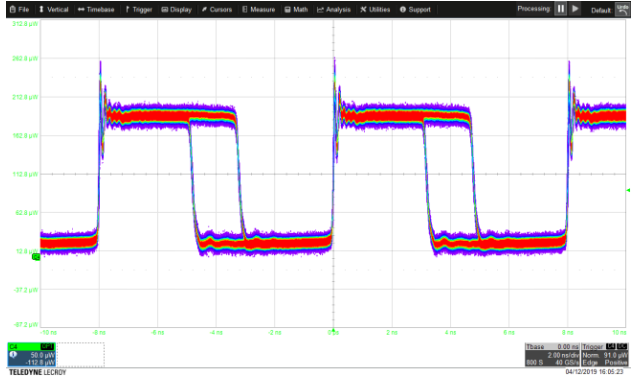

Fig. 10. A typical CDCM-20-1 signal obtained with the test transmitter.

The tester can emit a 62.5 MHz or 125 MHz clock with fixed or modulated duty cycle from 0% to ±45% around a mean 50% (10 settings). The modulation is governed by the transmitted data which are chosen from constant values, alternating ones and zeroes, a pseudo-random pattern (PRBS15), etc.

### B. Receiver

For the purpose of R&D, a CDCM receiver is built on the Front End Mezzanine (FEM) card which is part of the readout electronics for the high angle Time Projection Chambers (TPCs) in the on-going upgrade of the T2K neutrino oscillation experiment [14]. Although this application has modest clock synchronization requirements – several ns of skew and 100 ps rms clock jitter between the 32 expected target nodes is tolerable – I took the opportunity of a new board design to incorporate the required flexibility to study different clock and data forwarding schemes. The part of the FEM that is relevant to this study is shown in Fig. 11.

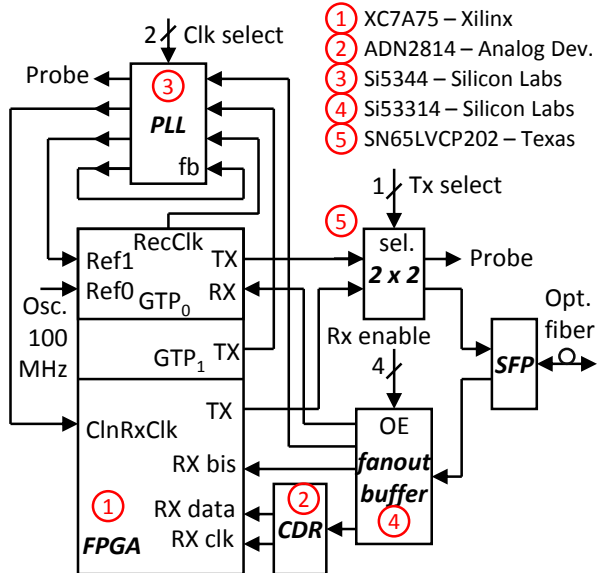

Fig. 11. Serial interface section of the FEM card. The optical link connects to back-end electronics that distributes a global clock, trigger and node specific configuration data in one direction, and gathers detector and monitoring data in the other direction.

The current baseline scheme for the upgrade of T2K is to use the external CDR chip in the receive direction and ordinary FPGA I/O pins to drive the SFP in the transmit direction. Protocols are detailed in [15]. Firmware changes allow the use of a GTP SerDes for RX only, TX only, or both. In these configurations, the external CDR chip is disabled, and the clock recovered by the receiver side of FPGA $GTP_0$ (RecClk) is cleaned by the external PLL to provide the reference clock (Ref1) for the transmit side. The configuration for CDCM studies takes two copies of the signals from the RX side of the optical transceiver. One copy is routed to an input of the external PLL and the other copy drives a pair of regular FPGA input pins. One output of the PLL is looped back for zero-delay operation. Ideally, a clock buffer should have been added in the feedback path to deskew the delay added by the clock fanout circuit on the main RX path. The transmit side of $GTP_1$ is connected to a clock input of the external PLL to study the behavior of the device under CDCM patterns.

## VI. EXPERIMENTAL STUDIES ON CLOCK-CENTRIC LINKS

### A. Receiver Based on a PLL Internal to the FPGA

In this test, the previously described CDCM-N-1 transmitter is connected to a FEM card via SFP transceivers and 4 m of optical fiber. This is schematically shown in Fig. 12.

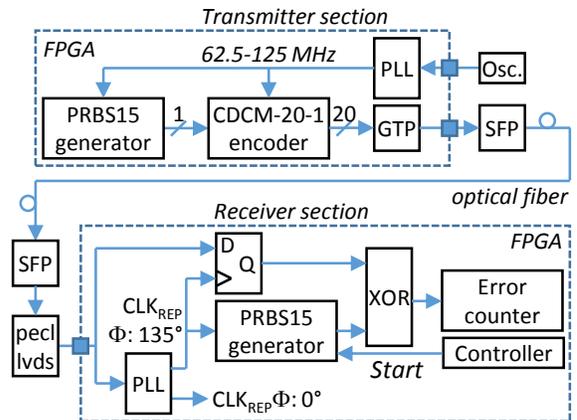

Fig. 12. Test implementation of a CDCM-3-1 transmitter and receiver using internal FPGA resources (target: Xilinx Artix 7).

A passive optical splitter allows probing transmitted signals with an 80 GSPS oscilloscope (LeCroy 820-ZiB) which is also used to measure various signals on the FEM card. After optical to electrical conversion on the FEM, received signals are routed to a pair of ordinary FPGA I/O pins (RX_bis on Fig. 11). Received signals are duplicated within the FPGA to feed an internal PLL and the Flip-Flop that captures serial data. One of the outputs of the PLL is set for 0° shift to reproduce the received clock while a distinct output generates the clock for data capture, at the same frequency as the forwarded clock in the present case. In theory, the clock for data capture should have a phase shift of 180° w.r.t. the forwarded clock. In practice, because of unequal delay in the clock and data paths, a static phase shift of 135° was found more adequate after a manual trial and error procedure. A digitally controlled delay (i.e. a Xilinx IDELAY2 primitive or the equivalent implemented in the fabric) could be inserted in the data path instead of performing a clock phase shift, but implementation was not attempted.

After data are extracted, they are compared to that produced by a local PRBS15 generator. A counter is incremented in case of mismatch between the received values and the expected ones. A finite state machine ensures synchronization of the local pseudo-random generator with the received pattern during an initialization phase. The input pre-divisor of the PLL is set to 4. This filters out the modulation of the input signal before the internal phase comparator of the PLL.

The receiver was successfully tested at 62.5 MHz / 62.5 Mbps and 125 MHz / 125 Mbps carrier clock frequency and user data rate respectively. At ±10% duty cycle modulation, no data transmission error were detected after over 8 hours of operation, corresponding to a BER inferior to $10^{-12}$. However, at ±5% modulation, data could not retrieved, probably because of the incorrect phase setting between the clock and data to capture. An automatic phase adjustment procedure would be required. The PLL locked successfully up to ±40% modulation but failed to lock at the highest modulation rate of ±45%. The random and deterministic jitters measured on the reproduced clock was 15 ps rms and 7 ps rms respectively at ±10% modulation. These figures were almost doubled at ±30% modulation. The histogram of delay between the optical signal probed at the optical fiber splitter and the reproduced clock at the receiver end is shown on Fig. 13.

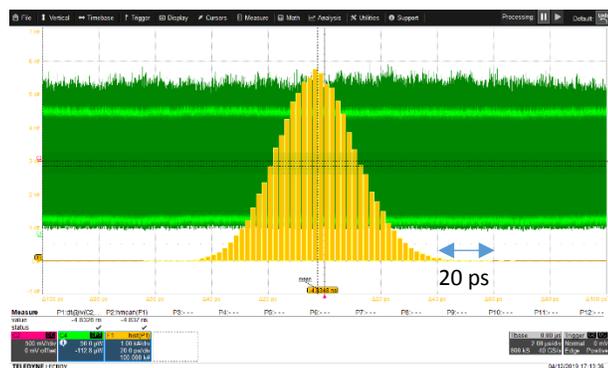

Fig. 13. Histogram of delay between the optical transmitted signal and the clock reproduced at the receiver end.

After ten consecutive receiver link re-configurations, it is observed that this mean delay remains constant within ~2 ps. This shows that no element with large unpredictable latency caused by a reset seems to be present.

These tests validate the basic concept of clock and data transmission using CDCM encoding and show how a synchronous serial data link with deterministic latency and clock transport capability can be built in an FPGA without using a dedicated high speed embedded SerDes. The proposed scheme can offer solutions for entry-level FPGA devices that do not have any high speed SerDes, or when these are used for some other purpose. Operation at a rate below the limit of such SerDes (e.g. 500 Mbps for Xilinx GTP transceiver) is possible. The structure can also easily be implemented in an ASIC.

### B. Clock Recovery with a Commercial Narrow-band PLL

In this test, one copy of the signals received through the optical transceiver of the FEM is input to the Si5344 PLL and a second copy is input to a regular I/O pin pair of the local FPGA (RX_bis on Fig. 11). The purified clock is probed at the output of the external PLL. Data are captured within the FPGA using an internally shifted version of the purified clock produced by the external PLL (ClnRxClk on Fig. 11). The datasheet of the Si5344 PLL specifies limits on the acceptable duty cycle for the input clock, 40%-60%, but no explanation is given on whether input clock duty cycle must remain fixed or if variations are acceptable. I have not found more precise specifications on this aspect in the documentation of other commercially available PLLs, probably because the use of these components with a voluntary modulated duty cycle of the reference clock has not been considered. The structure and operation of the pre-divisors of the Si5344 (and that of the other devices I have considered) are not explained in the datasheet, but it seems reasonable to assume that these are only sensitive to one edge of the input clock rather than both of them. Tests were done to assess the behavior of this PLL under the desired CDCM pattern.

The Si5344 is configured for zero-delay operation, with clock output #3 being fed-back to clock input #3. Silicon Laboratories recommends that the loop bandwidth is set to a minimum of 100 Hz minimum to guarantee a typical skew of 110 ps between the reference input and the feed-back input in zero-delay mode. The datasheet does not specify how stable this delay is, and which factors impact it.

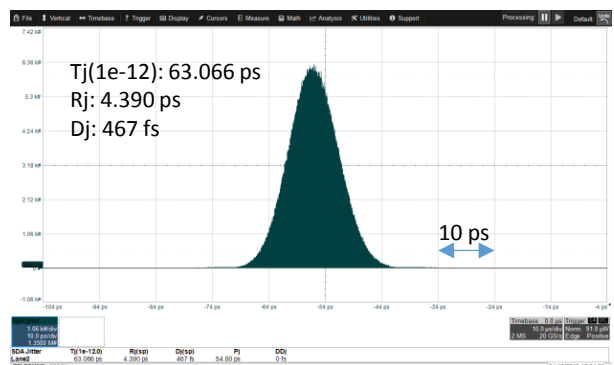

Fig. 14. Jitter of the 125 MHz clock received from a CDCM-20-1 transmitter with ±10% modulation after jitter filtering by a Si5444 PLL. Transported data is PRBS15 pattern at 125 Mbps.

Fig. 14 shows the jitter of the clock received from the test CDCM-20-1 transmitter (125 MHz carrier clock, ±10% duty cycle modulation around 50% by a PRBS15 bit stream at 125 Mbps) after jitter filtering by a Si5344 PLL. The measured random and deterministic jitters are 4.39 ps rms and 467 fs rms respectively. No significant change is found when Manchester data pre-encoding is enabled, and even when the transmitter is set to 0% duty cycle modulation. This suggests that the degradations induced by the modulation of the clock input to the PLL produce a negligible jitter transfer to the output.

### C. CDCM Fanout Demonstrator Board

This section details the implementation and performance evaluation of a CDCM fanout demonstrator board with one master port and two slave ports.

#### 1) Design

The simplified schematic of the CDCM fanout demonstrator board is shown in Fig. 15. The master SFP receiver is followed by a dual 1→2 clock buffer (Silicon Laboratories Si53308) which produces two copies of the received CDCM stream. One copy is fed to a jitter filter PLL (Silicon Laboratories Si5344) while the second copy is fed to a discrete D flip-flop (ON semiconductor MC100EP451). The Si5344 PLL is configured for zero-delay mode by looping back output $O_3$ to $IN_3$. The connection goes through the other half of the Si53308 clock buffer to compensate for the delay of this buffer. A 1→8 fanout chip (Texas Instruments CDCLVP1208) is placed at the Q output of the flip-flop. Two outputs of the octal fanout buffer drive the TX side of the two slave SFP ports of the board.

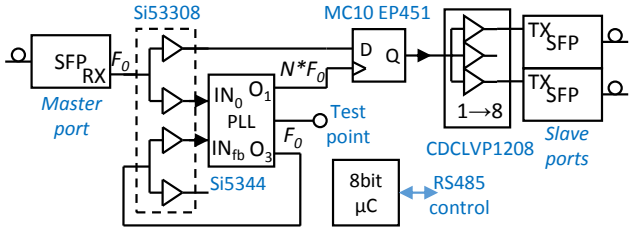

Fig. 15. Simplified schematic of the 1 to 2 CDCM fanout demonstrator board.

Output $O_1$ of the Si5344 PLL is connected to the clock input of the D flip-flop crossing the positive and negative sides of the differential pair. This ensures that the D flip-flop captures the CDCM stream at the optimum time which is on the falling edge of the extracted reference clock multiplied by the required factor. An 8-bit microcontroller is used for configuring the PLL. The board is also controllable from a PC via an RS-485 interface. It should be noted that this board does not contain any FPGA. The CDCM fanout is essentially composed of three non-programmable logic components surrounding a PLL.

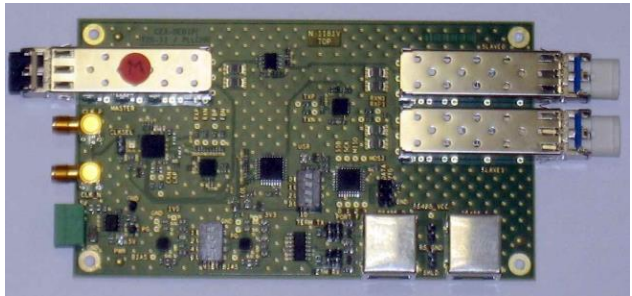

Fig. 16. Photograph of the CDCM fanout demonstrator board.

The transported serial data cannot by modified locally by this fanout board. Assuming that a CDCM-3-1 stream is received by the master port, the frequency setting of the output of the PLL connected to the D flip-flop ($O_1$) determines the function performed by the board as follows. When the frequency of PLL output $O_1$ is set to 3 times the frequency of the input CDCM carrier, the board becomes a repeater: the slave ports produce two jitter attenuated copies of the CDCM stream received by the master port. When the frequency of PLL output $O_1$ is set to be equal to the frequency of the carrier clock, the board becomes a serial data extractor terminal node: the slave ports output two copies of the transported serial data. Those bit streams no longer explicitly contains the carrier clock. The data extractor function is only available for CDCM-N-1 schemes because data extraction for other CDCM schemes requires additional logic that is not included in this voluntary simple demonstrator board. A photograph of the CDCM fanout demonstrator board is shown in Fig. 16. The optical transceivers, the PLL and the various PECL circuits, with their resistor terminations, are the main contributors to a rather high power consumption of 1.5 A.

*2) Performance evaluation*

Five of the above demonstrator board have been assembled. This allows testing at least two basic topologies: a chain of cascaded boards and a two-stage tree topology with one "trunk" node, two intermediate "branch" nodes, and two "leaf" nodes. Tested topologies are shown in Fig. 17.

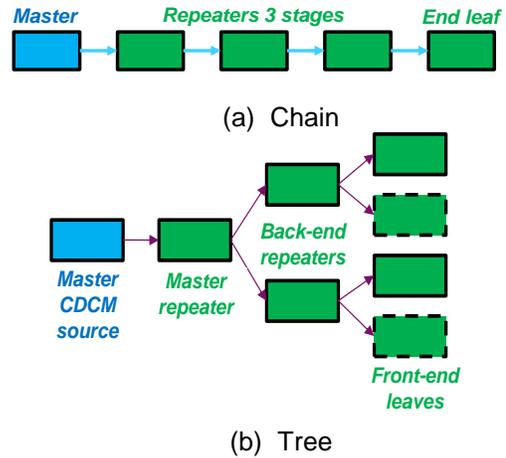

Fig. 17. Tested topologies.

The CDCM generator described in section V.A was used to provide the required stimuli. It was configured to carry a 125 MHz clock and 125 Mbps of data (PRBS15). Clock duty cycle modulation was set to ±10% around the 50% mean. It should be observed that after the first hop in the chain topology, or after the trunk-node in the tree topology, the CDCM-3-1 stream has a modulation of ±17% around a 50% mean, owing to the structure of the fanout demonstrator board and PLL settings.

Fig. 18 shows the jitter of the purified clock measured at the 4th hop in the cascaded chain. The measured random and deterministic jitters are 1.23 ps rms and 234 fs respectively, which is remarkably low. Nevertheless, even lower figures could probably be reached as the datasheet of the Si5344 PLL quotes 100 fs of rms jitter from 12 kHz to 20 MHz.

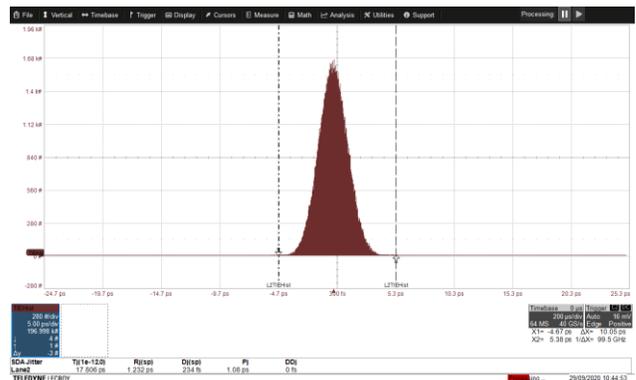

Fig. 18. Jitter of the extracted clock at the 4th hop in cascaded mode.

Fig. 19 shows the jitter obtained in the dual stage tree topology where the purified clock obtained at one leaf is used as the reference to measure the jitter of the purified clock obtained on the other leaf. The test equipment (LeCroy SDA 520ZiB) is used in serial data analyzer mode. This measurement is representative of the relative clock alignment at two distinct leaf end-points.

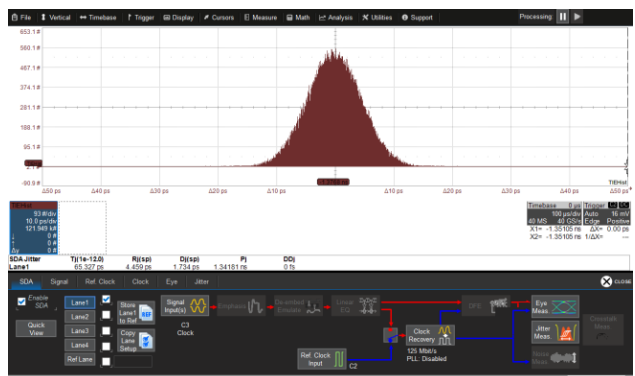
Fig. 19. Differential jitter on leaves in two stage tree mode.

The measured jitter is as low as ~5 ps rms (random jitter Rj=4.46 ps rms and deterministic jitter Dj=1.73 ps rms). For this measurement, a 14 GHz bandwidth differential probe was used on the leaf node that provided the reference clock, and a 2.5 GHz bandwidth unipolar probe was used on the second leaf. The measurement was repeated after setting the test equipment in oscilloscope mode. Two 2.5 GHz bandwidth unipolar probes were used. The histogram of the delay measured between the purified clocks at the leaf nodes was accumulated. This is shown on Fig. 20. The measured skew in this case was ~13 ps rms. Some of the observed degradation of performance is certainly caused by sub-optimal probing. Nonetheless, clock fidelity at the two end leaves remains high.

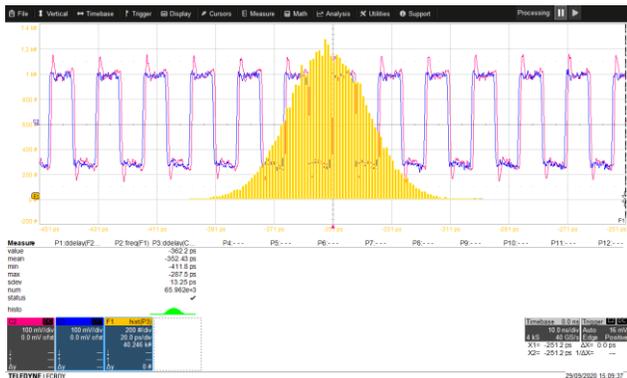
Fig. 20. Histogram of relative clock skew on two leaves in two stage tree.

The correctness of the extracted data was also checked at the output of each leaf node with a serial data analyzer (LeCroy SDA 6000) configured in bit error rate tester. No data error were revealed by this test after several tens of minutes of operation. Further tests over a longer period are planned to be done.

## VII. DISCUSSION

The proposed technique for clock and serial data encoding offers multiple advantages compared to ordinary methods. By design, it preserves entirely the timing content of the transported clock. Consequently, an ordinary PLL, narrow band and zero-delay if these features are also desired, can be used to purify directly the transported clock. This step does not rely on a CDR and it removes entirely the need to interpolate the serial clock edges which are missing when a traditional data encoding method is used. A CDCM receiver outputs the transported clock directly at the original frequency by placing a PLL at the closest place where the electrical signals from the physical link media are available. The circuitry for the capture of data is located downstream. This is in contrast with the common usage of an FPGA SerDes for clock steering where the serial clock for the capture of data is produced first, and the desired forwarded clock is obtained by frequency division, which has negative side effects on phase determinism and clock purity. With CDCM, the phase alignment and jitter of the forwarded clock can, in theory, be as good as what the selected receiving PLL can accomplish, i.e. few 100 fs clock jitter and 100 ps skew for today's most advanced industrial devices. CDCM serial links are FPGA vendor and model agnostic. Still, they can be built with embedded high speed FPGA SerDes, but they do not necessarily rely on them. This renders implementation possible with ordinary FPGA I/O pins and low-end devices. Design portability is improved. The minimal hardware complexity of CDCM-N-1 schemes also make them attractive for implementation in ASICs. Simulation models and software tools for tuning the adjustable parameters of a jitter cleaner PLL are normally provided by vendors along with their products. These facilitate system modelling and performance tuning. On the other hand, the jitter transfer function of an FPGA SerDes block is not easy to obtain and the settings of the embedded CDR (e.g. loop bandwidth) are not adjustable by the user.

One of the main limitations of the CDCM technique is low data throughput, which also leads to increased message latency. Implementation probably becomes very delicate when the frequency of the forwarded clock frequency is above several 100 MHz. Although most PLL models should be able to successfully lock on a CDCM stream, this must be verified in advance because the characteristics of a CDCM signal differ from those a PLL is normally designed for. At present only the Si5344 and Xilinx Artix 7 internal PLL have been tried. CDCM is media independent but precautions are needed when going optical: some Gigabit Ethernet SFP transceivers are capable of carrying a 125 MHz modulated clock, while others do not run properly with such input. Multi-rate transceivers with a sufficiently large operating speed range are a safe choice.

## VIII. CONCLUSIONS AND FUTURE WORK

In this paper I have proposed and started to investigate different implementations of an encoding technique for clock and data distribution over fast serial links where the overabundant data bandwidth is traded to make transmitters, receivers and fanout stages simpler, independent of any FPGA vendor and SerDes model, and potentially superior to mainstream solutions in terms of phase determinism and clock purity. While current techniques for clock distribution over serial links are indirect (clock recovery from the encoded data followed by frequency division), the proposed method is straightforward because the desired clock is effectively carried by the media. Serial data are embedded over the transported clock by modulating its duty cycle. I have illustrated this proposed concept of "clock-centric" serial communication by demonstrating the, apparently trivial, simultaneous transport of a 125 MHz reference clock and 125 Mbps stream of user data over a serial link where the receiver simply consists of a clock fanout chip, a PLL and elementary FPGA logic. The figures obtained on clock purity and latency determinism are within that reached by state-of-the art methods.

Future work will pursue the current R&D with the goal to prove or disprove some of the currently supposed merits of the proposed clock distribution approach.

## IX. QUESTIONS AND ANSWERS ON CDCM

*Q.1: "CDCM would introduce Data Dependent Jitter (DDJ) if data are simply encoded using pulse width modulation. 8B/10B encoding does not. Could you please comment on this?"*

What I did not clearly mention is that user data should be pre-encoded before they are embedded in a clock signal. This pre-encoding stage is required to keep the CDCM stream DC-balanced and to avoid generating data dependent jitter. The data pre-encoder will normally use an already known technique, such as Manchester encoding, 8B/10B encoding, or scrambling. However, the overhead of the data pre-encoder further reduces the usable bandwidth of a CDCM link, which is already low compared to traditional serial links. Hence, a data pre-encoding method that has no overhead, e.g. a self-synchronizing scrambler, can be preferable. Nonetheless, Manchester encoding could be attractive if losing half of the data throughput capacity is acceptable. Consider a CDCM-3-1 link with Manchester data pre-encoding. The serial stream on the media would be: $0, 1, D_i, 0, 1, \overline{D_t}, 0, 1, D_{i+1}, 0, 1, \overline{D_{t+1}}$, etc. The running disparity never exceeds one symbol and the stream is DC-balanced exactly after only six consecutive symbols. This will certainly contribute to minimize DDJ.

*Q.2: "Does CDCM logic have to run at a very fast clock rate, for example comparable to that of the internal of a GTP transceiver?"*

No, in most practical uses of CDCM, running at GHz clock rate will not be needed. The slowest possible rate for a CDCM transmitter (scheme CDCM-3-1) is 3 times the frequency of the carrier clock to be transported. The associated receiver only has to run at the nominal frequency of the carrier clock, not higher than that. For example, to transport a 125 MHz clock and 125 Mbps of data (DC-balanced), the logic of the corresponding CDCM-3-1 transmitter would need to run at 375 MHz. This can easily be reached with the output SerDes block of an ordinary FPGA I/O pin. The rate on the transport media would be 375 Mbaud. On the receiver side, the jitter filter PLL would run at 125 MHz, and the serial data capture logic would run at the same frequency.

From the point of view of total media bandwidth, the above example is equivalent to that required to transport clock and data over two distinct media: a plain 125 MHz clock takes a bandwidth equivalent to 250 Mbaud while a 125 Mbps data stream requires at least 125 Mbaud. This leads to a total required media capacity of 375 Mbaud. This is identical to the CDCM-3-1 scheme, except that CDCM takes the same physical media to carry both clock and data.

*Q.3: "How much of CDCM can you run just inside an FPGA without external components (using the PLL in the FPGA)?"*

The reception of several CDCM streams from independent clock sources would require one PLL per stream. Hence the number of PLLs inside the target FPGA would be the limiting factor. But CDCM is probably not the most adequate technique for this use-case. I think that CDCM may only be useful for the distribution of a common reference clock and synchronous data from a back-end node to one or multiple front-end nodes, like it is found in a typical data acquisition system. In the opposite link direction, from multiple front-end nodes to a commonly shared back-end node, using a traditional data transport technique seems more adequate because CDCM has poor data coding efficiency (40% maximum). When the transmitter of each front-end node is made synchronous to the common clock provided by the back-end node (using CDCM or any other technique), the back-end node can receive data from the front-end nodes using its own local reference clock, i.e. without the need of multiple CDR blocks, CDCM receivers or alike.

*Q.4: "In the test setup, an electrical fan was used for each FPGA. How much does the temperature of the FPGA affects clock accuracy (jitter or frequency)?"*

The CDCM fanout demonstrator board does not include any FPGA. As it is shown on the board schematic diagram, a CDCM fanout can be built only with a clock buffer, a PLL, discrete D flip-flops, etc. This structure copies the CDCM input stream on its output without modifying the transported data. An FPGA would only be needed if it is required to modify data on-the-fly. For a CDCM receiver that uses a PLL internal to an FPGA to purify the received clock, the accuracy in frequency and jitter will be determined by the characteristics of the device. Performance will be affected by process, temperature and supply voltage variations. I have not made such measurements. Xilinx Artix 7 datasheet does not give detail on these parameters. The maximum static phase offset of an internal PLL is quoted to be 120 ps. For a CDCM receiver based on a PLL external to the FPGA, performance will be determined by the characteristics of that PLL. The corresponding datasheets that I have seen do not give a sufficiently detailed characterization of the device. An interesting report on temperature dependence of several PLL models from Silicon Laboratories is found in [16].

*Q.5: "You have measured clock recovery jitter, but have you measured clock creation jitter caused by the FPGA encoding data?"*

That would be an interesting figure, but I have not made this measurement so far. In principle, a CDCM signal being a clock signal, its jitter should be measurable. And a CDCM signal being also a serial data stream, the jitter of the data section should be measurable too: eye opening, jitter histogram, bath tub diagram.

It should be noted that, in the architecture of the CDCM repeater I have shown, even if an FPGA is present, it will normally not contribute to the output jitter. Provided that a discrete re-timer D flip-flop clocked directly by the jitter purification PLL is used, only the jitter of the output of the PLL and the jitter added by the D flip-flop will matter. The additive jitter quoted for ON Semiconductor MC10EP451 differential D Flip-Flop is only 0.2 ps rms (typical) and 1 ps rms (maximum). For applications that would not require the lowest possible jitter, FPGA output pins could produce directly the final CDCM stream. In this case, the jitter caused by the serialization done by the FPGA would have to be considered. This remains to be quantified. If the jitter added by the FPGA remains acceptable, or if that jitter gets properly attenuated at the receiving end, the external re-timer D flip-flop stage that I have currently added would be superfluous. Removing it would simplify the design and lower power consumption and cost.